%% file: main.tex
\documentclass[sigconf]{acmart} 
\usepackage{multirow}
\usepackage{algorithm}
\usepackage{algpseudocode}
\usepackage{epstopdf}
\usepackage{array} 
\usepackage{hyperref}
\usepackage{float}
\usepackage{svg}
\usepackage{graphicx}
\usepackage{subcaption}
\usepackage{wrapfig}
\usepackage{xcolor}
\usepackage{makecell}
\usepackage{xcolor}

\usepackage[graphicx]{realboxes}

\copyrightyear{2025}
\acmYear{2025}
\setcopyright{cc}
\setcctype{by}
\acmConference[ICAIF '25]{6th ACM International Conference on AI in Finance}{November 15--18, 2025}{Singapore, Singapore}
\acmBooktitle{6th ACM International Conference on AI in Finance (ICAIF '25), November 15--18, 2025, Singapore, Singapore}\acmDOI{10.1145/3768292.3770427}
\acmISBN{979-8-4007-2220-2/2025/11}

\begin{document}

\input{sections/0_Front}
\input{sections/1_Introduction}
\input{sections/2_Related_Works}
\input{sections/3_Preliminaries}
\input{sections/4_Methodoloy}
\input{sections/5_Experiments}
\input{sections/6_Conclusion}

\bibliographystyle{ACM-Reference-Format}
\bibliography{ref}

\newpage
\appendix
\input{sections/Appendix}

\end{document}

%% file: sections/0_Front.tex
\title{Mean Variance Efficient Collaborative Filtering\\for Stock Recommendations}

\author{Munki Chung}
\authornote{These authors contributed equally to this work.}
\affiliation{%
  \institution{KAIST}
  \city{Daejeon}
  \country{Republic of Korea}
}
\email{moonki93@kaist.ac.kr}

\author{Junhyeong Lee}
\authornotemark[1]
\affiliation{%
  \institution{UNIST}
  \city{Ulsan}
  \country{Republic of Korea}
}
\email{jun.lee@unist.ac.kr}

\author{Yongjae Lee}
\authornote{Corresponding author}
\affiliation{%
  \institution{UNIST}
  \city{Ulsan}
  \country{Republic of Korea}
}
\email{yongjaelee@unist.ac.kr}

\author{Woo Chang Kim}
\authornotemark[2]
\affiliation{%
  \institution{KAIST}
  \city{Daejeon}
  \country{Republic of Korea}
}
\email{wkim@kaist.ac.kr}

\begin{abstract}
The rise of FinTech has transformed financial services online, yet stock recommender systems have received limited attention. Personalized stock recommendations can significantly impact customer engagement and satisfaction within the industry. However, traditional investment recommendations focus on high-return stocks or highly diversified portfolios, often neglecting user preferences. The former would result in unsuccessful investment because accurately predicting stock prices is almost impossible, whereas the latter would not be accepted by investors because many investors, including both individuals and institutional portfolio managers, who typically hold focused portfolios based on their investment strategies and interests. Collaborative filtering (CF) also may not be directly applicable to stock recommendations, because it is inappropriate to just recommend stocks that users like. The key is to optimally blend user’s preference with the portfolio theory. However, no existing model considers both aspects. We propose a simple yet effective model, called mean-variance efficient collaborative filtering (MVECF). Our model is designed to improve the Pareto optimality in a trade-off between the risk and return by systemically handling uncertainties in stock prices. Experiments on real-world data show our model can increase the mean-variance efficiency of recommended portfolios while sacrificing just a small amount of recommendation accuracy. Finally, we further show MVECF is easily applicable to the graph-based ranking model.
\end{abstract}

\keywords{Recommendation System, Mean-Variance Optimization}

\maketitle

%% file: sections/1_Introduction.tex
\section{Introduction}
\label{sec:intro}

Recommender systems have been one of the key success factors of companies who provide online services that they could attract more customers and make them spend more time in their services~\cite{Lu2015}. Numerous researchers have successfully applied recommender systems to relatively simple financial products such as bank products~\cite{Oyebode2020,Sharifihosseini2019}, insurances~\cite{Qazi2017,Rokach2013,Kanchinadam2018}, and real estate~\cite{Daly2014,Yuan2013}. The rise of FinTech has galvanized the participation of individual investors in stock markets~\cite{Hendershott2021}, but recommending stocks is not trivial due to the high uncertainty in their prices unlike other financial products. 

The necessity for stock recommender systems is in no doubt since investors are known to hold under-diversified portfolios~\cite{Dimmock2020}. The complexity of stock choices exacerbated by a flood of information and various market complexities often confuses individual investors. Companies offering precise and tailored stock recommendations consonant with an individual's existing portfolio will gain distinct customer loyalty. We list three key requirements for stock recommendation as a service. 

First, stock recommender systems \textbf{should account for \textit{investor’s preference.}} The traditional stock recommendation methods fall into two categories: 1) recommending stocks by predicting prices, and 2) recommending well-diversified portfolios constructed based on established portfolio theories. The first approach is inherently flawed due to the unpredictability of stock prices, largely attributed to market noise. The latter, while theoretically sound, often misaligns with a user's investment preference, and this is why investors still hold under-diversified portfolios even though there are a huge number of mutual funds managed by experts. Moreover, users who prefer direct investment experiences will not accept recommendations of funds. Note that most recommender systems for online services have been gaining great success from subtle but user-specific recommendations. Hence, stock recommendations should reflect users’ preferences that can be inferred from the users’ portfolios. 

Second, while the recommended stocks should align with the users' individual preferences, they simultaneously \textbf{should enhance the \textit{diversification} of their existing portfolio.} Direct applications of CF would probably make users’ portfolios more concentrated on some risk factors or industry sectors because recommended stocks would be similar to the users’ current holdings. More importantly, user preference does not necessarily lead to good investment. Therefore, stock recommendations should bring diversification benefits while satisfying the users’ preferences.

Third, they \textbf{should be proven in terms of \textit{ex-post performance} evaluations.} While wrong recommendations in the most of other online services would not severely harm users (e.g., waste of time, need to refund), wrong investment products would directly lead to monetary damages that cannot be simply returned. Hence, the performances of stock recommender systems should be carefully evaluated accounting for high uncertainties in stock prices. 

In this paper, we develop \textbf{a simple yet effective model for stock recommendation, called mean-variance efficient collaborative filtering (MVECF)}, that satisfies all three key requirements. We utilize the regularization technique to ensure that the recommendation is made based on the user’s current portfolio while increasing diversification effect by systemically handling uncertainties of stock returns, and the model is restructured to an ordinary weighted matrix factorization (WMF) form of Hu \textit{et al.}~\shortcite{Hu2008} to boost the computational efficiency. The experimental results show the proposed model improves the Pareto optimality in a trade-off between risk and return (i.e., the mean-variance efficiency). Moreover, the model outperforms baseline stock recommendation models in recommendation accuracy and exhibits superior investment performance than the conventional recommender systems in both in-sample and out-of-sample experiments. We further show that MVECF can be easily incorporated into graph-based ranking model by applying MVECF user-item ratings to the sampling process of the ranking systems.

%% file: sections/2_Related_Works.tex
\section{Related Works}
\subsection{Collaborative Filtering}
CF is the most popular approach in modern recommendation systems because of its efficiency in utilizing the entire user-item preference history. Hu \textit{et al.}~\shortcite{Hu2008} first developed a CF model for implicit feedback that predicts users' preferences by minimizing the square error of preference estimations. Another type of CF would be ranking models with the Bayesian personalized ranking (BPR) objective introduced by Rendle\textit{et al.}~\shortcite{Rendle2009}. Recently, graph-based ranking models like NGCF, LightGCN, UltraGCN, and HCCF~\cite{Wang2019,He2020,Mao2021,Xia2022} have achieved strong performance and gained widespread adoption for implicit feedback data.

Our work may look similar to a stream of research on the diversity and novelty of recommendations. These involve greedy re-ranking~\cite{Ziegler2005,Zhang2008,Vargas2014} and directly optimizing multi-objectives approaches~\cite{Su2013,Shi2012,Hurley2013,Wasilewski2016,Wasilewski2019}. However, the notion of recommendation diversity and novelty is quite different from portfolio diversification. While diversity in recommender systems focus on recommending dissimilar items, and novelty seeks items that are dissimilar to the items in the user’s past experiences, portfolio diversification considers the trade-off between minimizing risk and maximizing return of the entire portfolio. For example, if we define the dissimilarity measure as negative correlation between stock returns, novelty and diversity enhancing models may recommend stocks that reduce portfolio risk, but they might easily recommend poor stocks (e.g., negative returns) just because they have low correlation with the user's current holdings.

We incorporate LightGCN as a representative graph-based model and a model for enhancing novelty~\cite{Wasilewski2019} as baseline models.

\subsection{Stock Recommendations}
Previous studies on stock recommender systems can be categorized into three. First, some studies focus  on suggesting ways to measure similarities between stocks or equity funds and then apply existing CF methods ~\cite{Matsatsinis2009,Chang2018,DeRossi2020}. However, simply buying stocks that are similar to the current holdings or held by similar investors may make a portfolio more exposed to a certain type of risk (e.g., sector, style), contradicting the research goal of this work. Hence, we incorporate the portfolio theory to ensure proper diversification within the CF framework. 

Second, purely item-based recommendation. These studies try to find stocks that would have high returns in the future by analyzing item similarities between stocks~\cite{Yujun2016,Tu2016,Zheng2020,Gao2021}. This approach has a quite different perspective from recommender systems given that most methods in this category do not utilize user information, and giving the identical recommendation to all users will never yield even minimal recommendation accuracy. In addition, accurately predicting stock returns is almost impossible, because signals are dominated by noises in financial markets. Modern Portfolio Theory (MPT), established by Harry Markowitz~\shortcite{Markowitz1952}, can recommend an items to each user that maximizes the portfolio performance when added to the user's portfolio. We use this method as a baseline model for stock recommendation method in this category.

Third, recommendations based on user-item information and then diversification. For example, Musto \textit{et al.}~\shortcite{Musto2015} and Sweezy and Charron~\shortcite{Sweezy2018} recommend stocks based on the second approach and determine weights of stocks to reduce portfolio risk. However, adjusting portfolio weights after choosing stocks would have limited effects. On the other hand, Garc\'ia \textit{et al.}~\shortcite{Garcia2012} and Gonz\'alez~\shortcite{Gonzalez2012} carefully analyze user similarity and recommend a portfolio according to the user’s preference. This approach may be able to provide well-diversified portfolios to investors, but this has the same problem with existing mutual funds that investors prefer possessing a few stocks that they are interested in. In this category, the two step method of Sweezy and Charron~\shortcite{Sweezy2018} is included in the baselines.

%% file: sections/3_Preliminaries.tex
\section{Preliminaries}
\subsection{Weighted Matrix Factorization}
Consider $m$ users and $n$ stocks (items). Let vector $y_u \in \mathbb{R}^n$ represents the stock holdings of user $u$, and its components are the binary variable representing whether the user $u$ holds the stock $i$ is ${y_{ui}}$. In matrix form, the holdings of all users can be represented as the user-item interaction matrix ${\text{Y}} \in {\mathbb{R}^{m \times n}}$. WMF decomposes Y into the user embedding matrix ${\text{P}} \in {\mathbb{R}^{m \times l}}$ and the item embedding matrix $Q \in {\mathbb{R}^{n \times l}}$ with l number of latent factors as follows. 

\begin{align} \label{eq:wmf}
\min \mathop \sum \nolimits_{u,i} {c_{ui}}{\left( {{y_{ui}} - p_u^T{q_i}} \right)^2} + \lambda \mathop \sum \nolimits_u {p_u}^2 + \lambda \mathop \sum \nolimits_i {q_i}^2 
\end{align}

Here, ${y_{ui}} \in \mathbb{R}$ is an element of Y, and $p_u^T \in {\mathbb{R}^{1 \times l}}$ and $q_i^T \in {\mathbb{R}^{1 \times l}}\text{\;}$ are row vectors of P and Q, respectively. ${c_{ui}} \in \mathbb{R}$ is a hyperparameter indicating the confidence level about observation ${y_{ui}}$, which becomes large when ${y_{ui}} = 1$ and small when ${y_{ui}} = 0$, and $\lambda $ is a hyperparameter for ${{\text{L}}_2}$ regularization. ${\hat y_{ui}}$, the estimated preference of user u to item $i$ is $p_u^T{q_i}$. Hu \textit{et al.}~\shortcite{Hu2008} proposed an efficient alternating least squares (ALS) algorithm for solving Equation (\ref{eq:wmf}).

\subsection{Modern Portfolio Theory (MPT)}
Markowitz~\shortcite{Markowitz1952} was the first to mathematically define and analyze the risk and return of financial investments. The return of a risky asset was regarded as a random variable and the expected return was defined as its mean value and the risk was defined as its standard deviation. Then, $n$ risky assets can be described by their return vector $r \in {\mathbb{R}^n}$ with mean $\mu \in {\mathbb{R}^n}$ and covariance matrix $\Sigma \in {\mathbb{R}^{n \times n}}$. A portfolio of $n$ risky assets can be represented as a weight vector $w \in {\mathbb{R}^n}$, which should sum to one (i.e., $\mathop \sum \nolimits_i {w_i} = 1\text{\;}$), and it’s expected return and risk can be expressed as ${\mu ^T}w$ and $w^T\Sigma w$, respectively.

In Markowitz~\shortcite{Markowitz1952}, a convex quadratic programming problem (\ref{eq:MV}) was proposed to find a Pareto optimal portfolio between minimizing the risk and maximizing the expected return.

\begin{align} \label{eq:MV}
\mathop {\min }\limits_{\left\{ {w:\text{\;}\sum w = 1,w \geq 0} \right\}} \frac{\gamma }{2}w^T\Sigma w - {\mu ^T}w
\end{align}

It is called the mean-variance (MV) optimization, and the resulting optimal portfolios are called MV efficient portfolios. The term ‘efficient’ emphasizes the Pareto optimality of the solution from the trade-off between risk and return, where $\gamma$ is the parameter that represents the risk aversion of the investor. Markowitz~\shortcite{Markowitz1952} is the foundation of the MPT, and it is widely used in practice as well~\cite{Kim2021}.

An evaluation metric for measuring portfolio efficiency is proposed by Sharpe~\shortcite{Sharpe1966} which is called the Sharpe ratio. It is one of the most widely used performance measures in investment management. It is defined as a ratio between the expected return and the the standard deviation of portfolio returns as
\begin{equation} \label{eq:sr}
  SR\left( w \right) = \frac{{{\mu ^T}w}-r_f}{{\sqrt {{w^T}{\Sigma}w} }},
\end{equation}
where $r_f$ is the risk-free return. While the return of most secured investment products (e.g., 3-month U.S. T-bills) are used as a proxy for the risk-free return, we set it to 0 for simplicity.

It is worth noting that there are several studies that utilized the portfolio theory for information retrieval and recommender systems~\cite{Shi2012,Wasilewski2017,Wang2009,Wang2009_2}, however, their objective is to maximize the expected relevance of item ratings (mean of ${\hat{y}}_u$) and minimize the risk of relevance of item ratings (variance of ${\hat{y}}_u$) (or some of them maximize the variance to increase recommendation diversity). Clearly, they are focusing on the mean and variance of item ratings, not the financial context (mean and variance of stock returns).

%% file: sections/4_Methodoloy.tex
\section{Methodology}
The common goal of all recommendation systems is to recommend items that are likely to be selected by users in the future. Most recommender systems achieve this by recommending the top k items in the predicted ratings $\hat y$. In addition to that, the goal of this work is to generate $\hat y$ so that when the users accept the top k recommendations and add those items to their current portfolios ${y_u}$, the resulting portfolios will become more MV efficient (measured in the Sharpe ratio). Furthermore, our model should be able to adjust within two kinds of trade-off: 1) between recommendation performance and MV efficiency, 2) between portfolio risk and expected return. 

\subsection{Mean-Variance Regularization}
In this section, we develop the mean-variance efficient collaborative filtering (MVECF) method, which is a novel WMF model with regularization on the MV efficiency. In WMF model, the estimated item ratings (holdings) of user $u$ is ${\hat y_u} = Q{p_u}$, and we consider this as the user's final portfolio. The expected return and the risk of the portfolio would be written as ${\mu ^T}{\hat y_u}$ and $\hat y_u^T{\Sigma}{\hat y_u}$, respectively\footnote{Notice that $y_u$, which is a vector of binary variables, replaces the portfolio weight $w$ in Equation \ref{eq:MV}, which should sum to one. However, we do not normalize $\hat y_u$ for simplicity. Also, most of its elements will be continue to be less than 1 during training.}. We regularize `risk – return (i.e., $\frac{\gamma }{2}\hat y_u^T{\Sigma}{\hat y_u} - {\mu ^T}{\hat y_u}$)' of the user's portfolio, which is the objective function of the MV optimization problem in (\ref{eq:MV}), to the loss function of WMF. Then, the proposed model would recommend stocks based on CF while trying to minimize risk and maximize return. The resulting formulation is given in Equation (\ref{eq:mvecf_reg}). $\lambda_{MV}$ is a hyperparameter that controls the trade-off between the traditional recommendation performance and the MV efficiency of recommended portfolios.

\begin{align} \label{eq:mvecf_reg}
\begin{split}
 \mathop {\min }\limits_{P,Q} \mathop \sum \nolimits_{u,i} {c_{ui}}{\left( {{y_{ui}} - {{\hat y}_{ui}}} \right)^2} + \lambda \mathop \sum \nolimits_u {p_u}^2 + \lambda \mathop \sum \nolimits_i {q_i}^2 \\
 + {\lambda_{MV}}\mathop \sum \nolimits_u \left( {\frac{\gamma }{2}\hat y_u^T{\Sigma}{{\hat y}_u} - {\mu ^T}{{\hat y}_u}} \right) 
\end{split}
\end{align}

If we rewrite the last MV regularization term of Equation (\ref{eq:mvecf_reg}) in an elementwise expression with ${\mu_i}$ (elements of $\mu $), $\sigma_i^2$ and ${\sigma_{ij}}$ (diagonal and off-diagonal elements of ${\Sigma}$), Equation (\ref{eq:mvecf_reg}) can be rewritten as Equation (\ref{eq:mvecf_reg2}).
\begin{align} \label{eq:mvecf_reg2}
\begin{split}
\mathop {\min }\limits_{P,Q} \mathop \sum \nolimits_{u,i} \bigg[ {c_{ui}}&{{\left( {{y_{ui}} - {{\hat y}_{ui}}} \right)}^2} \\ 
+ {\lambda_{MV}}\frac{\gamma }{2}\big( \hat y_{ui}^2\sigma_i^2 & + {{\hat y}_{ui}}\mathop  \sum \nolimits_{j:j \ne i} {{\hat y}_{uj}}{\sigma_{ij}} \big) - {\lambda_{MV}}{{\hat y}_{ui}}{\mu_i} \bigg]\\
& + \lambda \mathop \sum \nolimits_u {p_u}^2 + \lambda \mathop \sum \nolimits_i {q_i}^2
\end{split}
\end{align}
We can see from Equation (\ref{eq:mvecf_reg2}) that the MV regularization would lower the rating of items with high variance (from $\hat y_{ui}^2\sigma_i^2$), lower the rating of items with high covariance with user holdings (from ${\hat y_{ui}}\mathop \sum \nolimits_{j:j \ne i} {\hat y_{uj}}{\sigma_{ij}}$), and raise the rating of items with high expected returns (from $ - {\hat y_{ui}}{\mu_i}$). Hence, it exactly delivers the desired effects. Equation (\ref{eq:mvecf_reg2}) can be solved with stochastic gradient descent (SGD) method.

\subsection{Restructuring MVECF into Ordinary WMF Form}
Here, we further increase the computational efficiency of MVECF by restructuring it into an ordinary WMF form (as in Equation (\ref{eq:wmf})) so that we can train it using the ALS algorithm developed by Hu \textit{et al.}~\shortcite{Hu2008}. ALS is known to converge much faster than SGD, and also SGD is highly sensitive to the choice of learning rate.
The trick is quite simple. In the MV regularization term in Equation (\ref{eq:mvecf_reg2}), we change ${\hat y_{ui}}\mathop \sum \nolimits_{j:j \ne i} {\hat y_{uj}}{\sigma_{ij}}$ into ${\hat y_{ui}}\mathop \sum \nolimits_{j:j \ne i} {y_{uj}}{\sigma_{ij}}/\left| {{y_u}} \right|$. Here, $\left| {{y_u}} \right|$ is the number of holdings of user $u$.

This trick has a nice theoretical property. Note that regularizing with ${\hat y_{ui}}\mathop \sum \nolimits_{j:j \ne i} {\hat y_{uj}}{\sigma_{ij}}$ would reduce the rating of items that have large covariances with ‘predicted’ user holdings. On the other hand, regularizing with the modified term ${\hat y_{ui}}\mathop \sum \nolimits_{j:j \ne i} {y_{uj}}{\sigma_{ij}}$ would reduce the rating of items that have large covariances with ‘true’ user holdings. Therefore, no matter which and how many items the user finally accepts from the recommended list, they would all possess diversification potential.
 
With the modification, the equation inside the summation over $u$ and $i$ in (\ref{eq:mvecf_reg2}) can be rewritten as
\begin{align}
\begin{split} \label{eq:mvecf_pre_wmf}
{c_{ui}}&{{\left( {{y_{ui}} - {{\hat y}_{ui}}} \right)}^2} + {\lambda_{MV}}\frac{\gamma }{2}\big( \hat y_{ui}^2\sigma_i^2 \\& + {\hat y_{ui}}\mathop \sum \nolimits_{j:j \ne i} {y_{uj}}{\sigma_{ij}}/\left| {{y_u}} \right| \big) - {\lambda_{MV}}{{\hat y}_{ui}}{\mu_i}.
\end{split}
\end{align}

We can see that both the first and the second term in (\ref{eq:mvecf_pre_wmf}) is a quadratic functions of ${\hat y_{ui}}$, and thus, we combine these two terms into one as in (\ref{eq:mvecf_pre_wmf2}).

\begin{align}
\begin{split} \label{eq:mvecf_pre_wmf2}
\big( {{c_{ui}} + \frac{\gamma }{2}{\lambda_{MV}}\sigma_i^2} \big)\hat y_{ui}^2 & \\ - \big( 2{c_{ui}}{y_{ui}} - \frac{\gamma }{2}{\lambda_{MV}}& \mathop \sum \nolimits_{j:j \ne i} \frac{{{y_{uj}}{\sigma_{ij}}}}{{\left| {{y_u}} \right|}} + {\lambda_{MV}}{\mu_i} \big){{\hat y}_{ui}} \\ + {c_{ui}}y_{ui}^2
\end{split}
\end{align}

If we let ${\tilde c_{ui}} = {c_{ui}} + \frac{\gamma }{2}{\lambda_{MV}}\sigma_i^2$ and define ${\tilde y_{ui}}$ to satisfy 
\[2{\tilde c_{ui}}{\tilde y_{ui}} = 2{c_{ui}}{y_{ui}} - \frac{\gamma }{2}{\lambda_{MV}}\mathop \sum \nolimits_{j:j \ne i} \frac{{{y_{uj}}}}{{\left| {{y_u}} \right|}}{\sigma_{ij}} + {\lambda_{MV}}{\mu_i}\]
then (\ref{eq:mvecf_pre_wmf2}) becomes 

\begin{align}\label{eq:perfect_square}
\begin{split}
&{\tilde c_{ui}}\hat y_{ui}^2 - 2{\tilde c_{ui}}{\tilde y_{ui}}{\hat y_{ui}} + {c_{ui}}y_{ui}^2\\
&={\tilde c_{ui}}{\left( {{{\tilde y}_{ui}} - {{\hat y}_{ui}}} \right)^2} - {\tilde c_{ui}}\tilde y_{ui}^2 + {c_{ui}}y_{ui}^2.
\end{split}
\end{align}
Since $ - {\tilde c_{ui}}\tilde y_{ui}^2 + {c_{ui}}y_{ui}^2$ is independent with P and Q, minimizing (\ref{eq:perfect_square}) is equivalent to minimizing ${\tilde c_{ui}}{\left( {{{\tilde y}_{ui}} - {{\hat y}_{ui}}} \right)^2}$. Thus, MVECF can be written as
\begin{align} \label{eq:mvecf_wmf}
  \min \mathop \sum \nolimits_{u,i} {\tilde c_{ui}}{\left( {{{\tilde y}_{ui}} - {{\hat y}_{ui}}} \right)^2} + \lambda \left( {\mathop \sum \nolimits_u {p_u}^2 + \mathop \sum \nolimits_i {q_i}^2} \right).
\end{align}
Note that (\ref{eq:mvecf_wmf}) has exactly the same form of ordinary WMF with modified ratings ${\tilde y_{ui}}$ and their weighting coefficients ${\tilde c_{ui}}$. To interpret ${\tilde y_{ui}}$ and ${\tilde c_{ui}}$, we define two MV related parameters $c_{ui}^{MV} = \frac{\gamma }{2}{\lambda_{MV}}\sigma_i^2$ and $y_{ui}^{MV} = \left( {\frac{{{\mu_i}}}{\gamma } - \frac{1}{2}\mathop \sum \nolimits_{j:j \ne i} \frac{{{y_{uj}}}}{{\left| {{y_u}} \right|}}{\sigma_{ij}}} \right)/\sigma_i^2$. Then the definition ${\tilde y_{ui}}$ can be rewritten as (\ref{eq:mvecf_y}), where ${\tilde c_{ui}} = {c_{ui}} + c_{ui}^{MV}$.
\begin{align} \label{eq:mvecf_y}
\begin{split}
{\tilde y_{ui}} & = \frac{{2{c_{ui}}{y_{ui}} + {\lambda_{MV}}\left( {{\mu_i} - \frac{\gamma }{2}\mathop \sum \nolimits_{j:j \ne i} \frac{{{y_{uj}}}}{{\left| {{y_u}} \right|}}{\sigma_{ij}}} \right)}}{{2{{\tilde c}_{ui}}}} \\
& = \frac{{{c_{ui}}{y_{ui}} + \frac{\gamma }{2}{\lambda_{MV}}\sigma_i^2\frac{{\frac{{{\mu_i}}}{\gamma } - \frac{1}{2}\mathop \sum \nolimits_{j:j \ne i} \frac{{{y_{uj}}}}{{\left| {{y_u}} \right|}}{\sigma_{ij}}}}{{\sigma_i^2}}{{\;}}}}{{{{\tilde c}_{ui}}}} \\
& = \frac{{{c_{ui}}{y_{ui}} + c_{ui}^{MV}y_{ui}^{MV}{{\;}}}}{{{{\tilde c}_{ui}}}}
\end{split}
\end{align}

We can see that the modified target rating ${\tilde y_{ui}}$ in Equation (\ref{eq:mvecf_y}) is a weighted sum of the user’s current holdings ${y_{ui}}$ and the MV rating $y_{ui}^{MV}$, where $y_{ui}^{MV}$ would have a large value when the mean return of item $i$ (${\mu_i}$) is high and the risk of item $i$ ($\sigma_i^2$) is low. Therefore, the recommendation ${\hat y_{ui}}$ would directly reflect the MV rating to favor items with better risk-return profiles. Also, the weighting term ${\tilde c_{ui}}$ becomes large when $\sigma_i^2$ is large, and thus, the model focuses more on matching ratings of risky items compared to safe items. By simply changing the true rating ${y_{ui}}$ into a weighted sum of ${y_{ui}}$ and $y_{ui}^{MV}$, we can train a WMF model to make user preferred recommendations, while making the resulting portfolio more efficient in terms of risk-return trade-off. The model can be easily trained by ALS.

%% file: sections/5_Experiments.tex
\section{Experiments}
\subsection{Data}
For our experiments we use the Survivorship-Bias-Free US Mutual Fund dataset from the Center for Research in Security Prices (CRSP), retrieved via Wharton Research Data Services. This dataset provides monthly snapshots of each mutual fund's holdings from January 2002 through December 2024.

\begin{figure}[t]
\centering
\includegraphics[width=1\columnwidth]{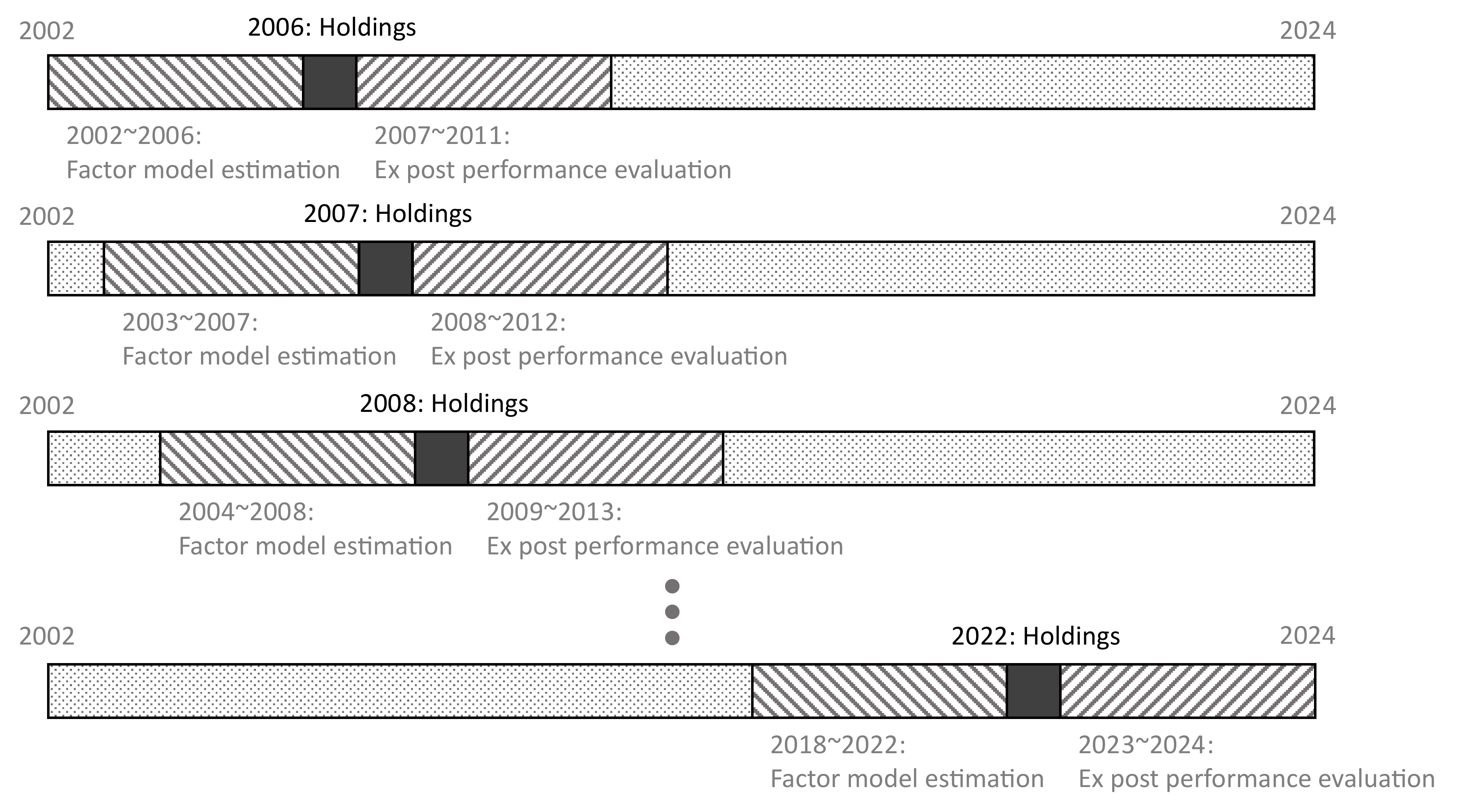}
\caption{Yearly sub-dataset construction}
\label{fig:data}
\end{figure}

\begin{table}[t]
\caption{\label{table:crsp_extended} Data description of 17 sub-datasets from CRSP (2006-2022). $m$ is number of users, $n$ is number of items and $\bar{h}$ is the average number of item holdings of users.}
\centering
\footnotesize
\setlength{\tabcolsep}{3.5pt}
\begin{tabular}{lccccccccc}
\toprule
Year & 2006 & 2007 & 2008 & 2009 & 2010 & 2011 & 2012 & 2013 & 2014 \\
\midrule
$m$ & 1900 & 459 & 2054 & 3332 & 3354 & 3553 & 3801 & 3776 & 3737 \\
$n$ & 785 & 603 & 758 & 774 & 764 & 813 & 873 & 982 & 1115 \\
$\bar{h}$ & 44.07 & 41.74 & 42.70 & 48.38 & 47.02 & 52.12 & 53.13 & 55.75 & 55.76 \\
\midrule
Year & 2015 & 2016 & 2017 & 2018 & 2019 & 2020 & 2021 & 2022 & \textbf{Avg.} \\
\midrule
$m$ & 3884 & 3906 & 4001 & 4409 & 4592 & 4696 & 4556 & 4819 & \textbf{3578} \\
$n$ & 1236 & 1324 & 1399 & 1471 & 1609 & 1709 & 1882 & 2210 & \textbf{1136} \\
$\bar{h}$ & 59.13 & 61.37 & 65.02 & 69.73 & 76.07 & 79.43 & 83.52 & 88.51 & \textbf{60.20} \\
\bottomrule
\end{tabular}
\end{table}

We split the dataset into yearly sub-datasets as shown in Figure \ref{fig:data}. In our experiment, we recommend items to users at the end of each year, so we use holdings snapshot reported in December of year T as the holdings of year T. We define preferred items as the current holdings to avoid recommending the items that were recently sold by the user. Although we have monthly holdings data, portfolios do not change much every month, so we create yearly sub-datasets for more robust experiments.

For MVECF model, we need stock mean returns and covariance matrix should be estimated before the recommendation. Hence, we estimate them using returns data in the past 5 years (years T-4 to T). For ex post performance evaluation, we use the next 5 years returns data (years T+1 to T+5). For example, when we are running an experiment for the year 2015 dataset, the user-item interaction data is constructed from the holdings in December 2015, the stock mean return and covariance matrix are estimated with the stock returns data from January 2011 to December 2015, and the stock returns data from January 2016 to December 2020 is used for ex-post performance evaluation.

As a result, we get 17 yearly sub-datasets from 2006 to 2022. The number of users and items, and the average number of holdings in resulting 17 yearly sub-datasets are summarized in Table \ref{table:crsp_extended}. The user-item interaction data in each dataset is divided into train, test, and validation data at a ratio of 8:1:1.

\begin{table*}[t]
\caption{Average evaluation metrics across all 17 datasets for recommended portfolios by MVECF models with different hyperparameter values. Higher $\lambda_{MV}$ values improve portfolio efficiency (e.g., $\Delta\text{SR}$ increases from 0.0182 to 0.0724) at the cost of recommendation accuracy (MAP@20 decreases from 0.2272 to 0.2112). The risk aversion parameter $\gamma$ controls the risk-return trade-off, with higher values reducing portfolio risk more aggressively while maintaining over 99\% improvement rate in SR.}
\centering
\begin{tabular}{@{\hspace{0.5em}}c@{\hspace{1em}}c@{\hspace{1.5em}}c@{\hspace{1.5em}}c@{\hspace{1.5em}}c@{\hspace{1.5em}}c@{\hspace{1.5em}}c@{\hspace{1.5em}}c@{\hspace{0.5em}}}
\toprule
& \textbf{Parameters} & \multicolumn{6}{c}{\textbf{Performance metrics}} \\
\midrule
& $\lambda_{\text{MV}}$ & $\Delta\mu$ & $\Delta\sigma$ & $\Delta$SR & P(SR $>$ SR$_{\text{init}}$) & MAP@20 & Recall@20 \\
\midrule
MVECF$_{\text{wmf}}$ & 0.1  & 0.0003  & -0.0068 & 0.0182 & 0.7276 & 0.2272 & 0.8675 \\
                     & 1    & 0.0005  & -0.0104 & 0.0277 & 0.8425 & 0.2268 & 0.8646 \\
                     & 10   & 0.0013  & -0.0245 & 0.0724 & 0.9921 & 0.2112 & 0.7832 \\
\midrule
& $\gamma$ & $\Delta\mu$ & $\Delta\sigma$ & $\Delta$SR & P(SR $>$ SR$_{\text{init}}$) & MAP@20 & Recall@20 \\
\midrule
MVECF$_{\text{wmf}}$ & 1    & 0.0083  & -0.0135 & 0.0712 & 0.9908 & 0.2118 & 0.7837 \\
                     & 3    & 0.0013  & -0.0245 & 0.0724 & 0.9921 & 0.2112 & 0.7832 \\
                     & 5    & -0.0008 & -0.0283 & 0.0745 & 0.9912 & 0.2097 & 0.7796 \\
\bottomrule
\end{tabular}
\label{table:mvecf_wmf_performance}
\end{table*}

\subsection{Models}
We consider the WMF form version of MVECF ($\text{MVECF}_\text{wmf}$) in Equation (\ref{eq:mvecf_wmf}). For baseline models, we use two simple baselines (WMF and BPR) and one graph-based recommender system, LightGCN. In addition, we consider three categories of models for stock recommendations as baseline models to enhance MV efficiency of recommended portfolio with existing methods. First, the novelty enhancing BPR model ($\text{BPR}_\text{nov}$) of Wasilewski \textit{et al.}~\shortcite{Wasilewski2019}. For this model, we define the dissimilarity (distance) between two items $i$ and $j$ as $\sqrt{1 - \rho_{ij}}$, where $\rho_{ij}$ is the return correlation between $i$ and $j$. Second, the 2-step method of Sweezy and Charron~\shortcite{Sweezy2018} introduced in Section 2.1. It filters K items using a base recommendation model with a cutoff K, and then makes the final recommendation by re-ranking the scores of top-k items using the MPT method. We consider the version with WMF as the base recommendation method ($\text{2Step}_\text{wmf}$). Lastly, the method incorporating MPT that recommends items that maximize Sharpe ratio of recommended portfolio. We denote this method as $\text{MPT}_\text{topSR}$. In short, we compare $\text{MVECF}_\text{WMF}$ with six baseline models: WMF, BPR, LightGCN, $\text{BPR}_\text{nov}$, $\text{2Step}_\text{wmf}$, and $\text{MPT}_\text{topSR}$.

\begin{figure*}[t!]
\centering
\includegraphics[width=2\columnwidth]{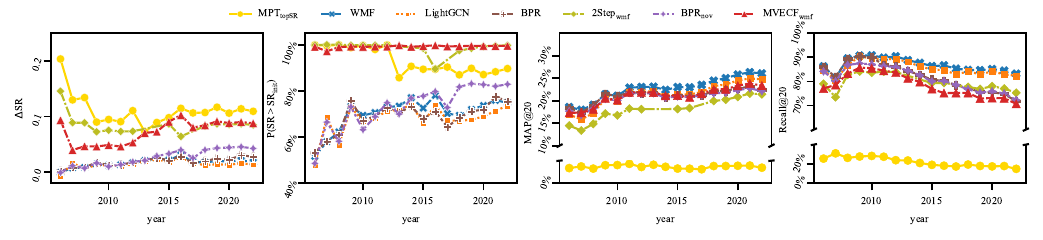}
\caption{Performance comparison between MVECF and baseline models across 17 yearly datasets (2006-2022). MVECF$_{\text{wmf}}$ consistently achieves superior portfolio efficiency with $\Delta\text{SR}$ around 0.1 and nearly 100\% SR improvement rate, while maintaining competitive recommendation accuracy (MAP@20 $\approx$ 20\%, Recall@20 $\approx$ 80\%). Traditional recommenders (WMF, LightGCN, BPR) show minimal portfolio improvements, while MPT$_{\text{topSR}}$ exhibits high variance across years.}
\label{fig:main}
\end{figure*}

\begin{figure*}[t!]
\centering
\includegraphics[width=1.2\columnwidth]{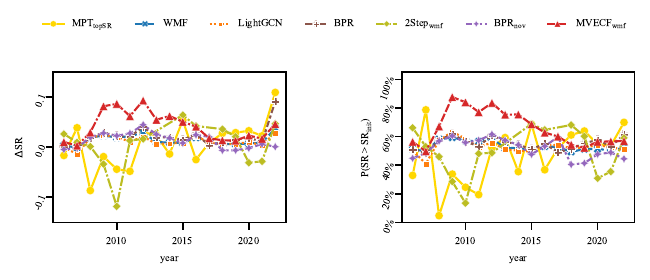}
\caption{Ex-post performance comparison using 5-year realized returns. MVECF$_{\text{wmf}}$ demonstrates robust out-of-sample performance with positive $\Delta\text{SR}$ and dominating SR improvement rate across most years. In contrast, MPT$_{\text{topSR}}$ and 2Step$_{\text{wmf}}$ show unstable performance with frequent Sharpe ratio deterioration.}
\label{fig:ex_post}
\end{figure*}

\subsection{Performance Evaluation}
The performance of stock recommender systems should be evaluated in two aspects: 

\noindent
1) Whether users will like the recommended stocks

\noindent
2) Whether the recommendations will improve the users' portfolios

The first one can be measured based on conventional recommendation performance (precision, recall). To be more specific, top 20 items of each user's test data are recommended and mean average precision at 20 (MAP@20) and recall at 20 (Recall@20) are used for conventional recommendation performance evaluation.

The second one should be based on MV efficiency. For this, we recommend top 20 items of all non-holding items to users. This is because recommending items only in pre-chosen test set is inappropriate for investment setting. We calculate the Sharpe ratios (equation (\ref{eq:sr})) of the initial portfolio (before recommendation) and the recommended portfolio (after adding 20 recommended items). Both portfolios are constructed as equally weighted portfolios because MVECF models recommend just items (stocks), not their weights.

Next, we use two measures that are based on the calculated Sharpe ratios. First, ${\Delta \text{SR}}=\text{SR}-\text{SR}_\text{init}$, the improvement in the Sharpe ratio, where SR and $\text{SR}_\text{init}$ are Sharpe ratios of recommended and initial portfolios, respectively. Second, $\text{P}(\text{SR} > \text{SR}_\text{init})$, the percentage of users whose Sharpe ratio has increased after recommendation. While ${\Delta \text{SR}}$ would show the amount of improvement, $\text{P}\big( \text{SR} > \text{SR}_\text{init} \big)$ would show how many users get the improvement.

As noted in Introduction, ex-post evaluation is particularly important in investment management. Hence, the same metrics, ${\Delta \text{SR}}$ and $\text{P}(\text{SR} > \text{SR}_\text{init})$, are calculated with 5 years realized return and risk of the recommended portfolio using Equation (\ref{eq:sr}). Note that Sharpe ratio is undoubtedly the most widely used investment performance measure in both practice and academia.

We also analyze the improvement in mean return ${\Delta\mu}$ and risk ${\Delta\sigma}$ of recommended portfolio to see the risk-return tradeoff of MVECF. The hyperparameters are tuned within the validation set (10\% of total data). Experiments are repeated with different values of balancing hyperparameter ${{\lambda}_{\text{MV}}}$ in 0.1, 1, 10 and risk-aversion level ${\gamma}$ in 1, 3, 5. 

\subsection{Experimental Results}
Before we demonstrate the relative performances of MVECF models to baseline models, let us check whether MVECF performs as we expected. Table \ref{table:mvecf_wmf_performance} shows the averages of various evaluation metrics of MVECF$_{\text{WMF}}$ with different values of $\lambda_\text{MV}$ and $\gamma$. The upper half shows the results of various values of $\lambda_\text{MV}$ when $\gamma$ is fixed to 3. The model shows better $\Delta \text{SR}$ and $\text{P}\left( \text{SR} > \text{SR}_{\text{init}} \right)$ when $\lambda_{MV}$ is large, and better MAP@20 and Recall@20 when $\lambda_{MV}$ is small. We can easily see that the tradeoff between MV efficiency and conventional recommendation performance is well controlled by $\lambda_{MV}$. 

The bottom half is the results of varying $\gamma$ when $\lambda_{MV}$ is fixed to 10. As we assume a more risk-averse user by increasing $\gamma$, it clearly shows the MVECF$_{\text{wmf}}$ model focuses more on reducing risk than increasing return. Hence, $\gamma$ controls the risk-return tradeoff as desired. We argue that the results are consistent across all 17 sub-datasets.

Now, we compare the MV efficiency and recommendation performances of MVECF with baseline models. Figure \ref{fig:main} shows the results of conventional recommender systems, existing stock recommendation models, and MVECF$_{\text{wmf}}$ with $\lambda_{MV} = 10$ and $\gamma = 3$ for a total of 17 sub-datasets. Each marker in the graph represents the average performance of all users in one yearly sub-dataset. Figure \ref{fig:ex_post} represents the ex-post Sharpe ratio performance.

From the two figures, it is evident that $\text{MVECF}_\text{wmf}$ outperforms conventional recommender systems in terms of MV efficiency (i.e., $\Delta \text{SR}$ and $\text{P}(\text{SR} > \text{SR}_{\text{init}})$). $\text{MVECF}_\text{wmf}$ exhibits dominating performance with $\text{P}(\text{SR} > \text{SR}_{\text{init}})$ values near 100\% for all datasets, implying that MVECF is beneficial for almost all investors in improving MV efficiency. Conventional recommender systems, however, exhibit $\Delta \text{SR}$ and $\text{P}(\text{SR} > \text{SR}_{\text{init}})$ values not much different from each other, as they do not address portfolio diversifications.

Regarding recommendation performance, $\text{MVECF}_\text{wmf}$ naturally shows inferior performance compared to conventional recommendation systems. However, the difference is not substantial. The average decrease in recommendation performance across 17 sub-datasets compared to the best performing LightGCN is less than 5\% in MAP@20 and less than 10\% in Recall@20. This is relatively small compared to the improvement in MV efficiency.

As for existing stock recommendation models, Figure \ref{fig:main} indicates that two-step method $\text{2Step}_{\text{wmf}}$ achieves similar MV efficiency levels to $\text{MVECF}_\text{wmf}$ and $\text{MPT}_\text{topSR}$ shows greater $\Delta \text{SR}$, whereas $\text{BPR}_{\text{nov}}$ does not. However, a clear difference can be seen from the ex-post performance in Figure \ref{fig:ex_post}. While $\text{MVECF}_\text{wmf}$ outperforms all other models in most cases, the two-step model and $\text{MPT}_\text{topSR}$ show lower $\Delta \text{SR}$ than all other models, even negative, and $\text{P}(\text{SR} > \text{SR}_{\text{init}})$ below 50\% in most of the sub-datasets. This suggests that these methods may demonstrate a sound MV efficiency in theory, but the actual investment performance can be significantly deficient. Finally, it is obvious that recommendation performances of $\text{MPT}_\text{topSR}$ is far below those of all other models since it does not consider the user-item information, and also two-step models, especially $\text{2Step}_{\text{wmf}}$ exhibits lower MAP@20 than $\text{MVECF}_\text{wmf}$ across all sub-datasets and show similar levels of Recall@20.

\subsection{Incorporating MVECF into Ranking Models}
Most of the recent state-of-the-art models on recommender systems are based on ranking models with implicit feedbacks. Note that MVECF is developed based on WMF, which is a rating-prediction model. However, the key point of MVECF is to modify the true rating ${y_{ui}}$ to ${\tilde y_{ui}}$, which incorporates the MV efficiency of items. Using this, we show that MVECF can be easily incorporated into the ranking models.

In traditional ranking models, items associated with real user-item interactions are considered as positive samples, while negative samples are chosen from the items without such interactions. We propose an MV efficient sampling scheme that identifies items with ${\tilde y_{ui}} > \tau $ as positive samples and items with ${\tilde y_{ui}} < \tau $ as negative samples, where $\tau $ is a predefined threshold level.

As discussed in Section 4.2, ${\tilde y_{ui}}$ is a weighted sum of the true rating ${y_{ui}}$ and the MV rating $y_{ui}^{MV}$. The MV rating $y_{ui}^{MV}$ is bad (good) when the item is highly (less) correlated with the user's portfolio and/or has low (high) expected return. Hence, even though a user actually holds item $i$, if $y_{ui}^{MV}$ is really bad, it will be classified as a negative sample in MV efficient sampling. Similarly, items with really good $y_{ui}^{MV}$ will be regarded as positive samples, regardless of actual interactions ${y_{ui}}$. Given the sparsity of true positive items, we set the threshold $\tau $ so that 1\% of the original negative samples can be converted to positive samples in MV efficient sampling. 

Table \ref{table:GCN} presents the MV efficiency and recommendation performance of LightGCN, which exhibited the best recommendation performance in Section 5.4, as well as LightGCN with the MV efficient sampling. The results clearly show that the state-of-the-art graph based ranking models can be easily extended to improve MV efficiency (${\Delta \text{SR}}$ and ${\text{P}}({\text{SR}} > {\text{S}}{{\text{R}}_{{\text{init}}}})$) with simple modification through the MV efficient sampling scheme.

\begin{table}[t]
\caption{\label{table:GCN} Average of evaluation metrics across all 17 datasets for LightGCN and the model with MV efficient sampling $\text{LightGCN}_\text{MVECF}$.}
\centering
\scalebox{0.9}{
\begin{tabular}{ccccc}
\toprule
 & ${\Delta \text{SR}}$ & ${\text{P}}({\text{SR}} >  {\text{S}}{{\text{R}}_{{\text{init}}}})$ & MAP@20 & Recall@20 \\ 
 \midrule
LightGCN            & 0.0139 & 0.6785 & 0.2135 & 0.8549 \\
$\text{LightGCN}_\text{MVECF}$ & 0.0877 & 0.9861 & 0.1468 & 0.6170 \\
\bottomrule
\end{tabular}%
}
\end{table}

%% file: sections/6_Conclusion.tex
\section{Conclusion}
In this paper, we proposed the mean-variance efficient collaborative filtering (MVECF) for stock recommendation that can systemically handle the risk-return profile of recommended portfolios while recommending stocks with the consideration of user preferences. Starting from a simple regularization, we were able to derive MVECF as an ordinary WMF form. The performances of portfolios recommended by MVECF outperformed other recommender systems in both in-sample and out-of-sample settings with only minimal reductions in the recommendation performance. Furthermore, we demonstrated that the modified user-item rating of MVECF can be integrated into the positive and negative sampling of ranking models, allowing graph-based models to offer MV efficient recommendations for users. 

The importance of this research lies in addressing the unique challenges of stock recommendation within the rapidly evolving fintech industry. As personalized stock recommendations become increasingly relevant for attracting and retaining customers, our approach can significantly enhance customer engagement and satisfaction, providing investment companies and online brokers with a competitive edge in the fintech landscape. This is the first study to identify the key requirements for stock recommender systems and develop a proper CF model for stock recommendations to fully utilize the user-item interactions. We believe that our study can encourage many researchers to develop more advanced stock recommender systems that can properly handle the risk-return characteristics of stocks as well as the preference of users.

\begin{acks}
This work was supported by the National Research Foundation of Korea (NRF) grant funded by the Korea government (MSIT) (No. NRF-2022R1I1A4069163).
\end{acks}

%% file: sections/Appendix.tex
\section{MPT stock recommendation}
As we discussed in the section 2.2, We included a baseline model which can recommend items that maximizes the portfolio performance by incorporating MPT. Given that means and variances of two assets $s_1$ and $s_2$ are $\mu_1$, $\mu_2$ and $\sigma_1^2$ and $\sigma_2^2$, and their covariance is $\sigma_{12}$, then the mean and the variance of a portfolio with weight $w_1$ for $s_1$ and $w_2$ for $s_2$ are $\mu_1w_1 + \mu_2w_2$ and $\sigma_1^2w_1^2 + \sigma_2^2w_2^2 + 2w_1w_2\sigma_{12}$. The Sharpe ratio for the portfolio with the two assets is as follow
\begin{equation} \label{eq:sr_appendix}
  \frac{\mu_1w_1 + \mu_2w_2-r_f}{\sqrt{\sigma_1^2w_1^2 + \sigma_2^2w_2^2 + 2w_1w_2\sigma_{12}}}
\end{equation}
If the asset $s_1$ is the current holdings of a user $u$ and $s_2$ is an additional item to be recommended, since we assumed equal weight portfolio for simplicity, $w_1$ becomes $\frac{|y_u|}{|y_u|+1}$ and $w_2$ becomes $\frac{1}{|y_u|+1}$ where $|y_u|$ is number of holdings of the user, and lastly $\sigma_{12}=\sigma_{1}\sigma_{2}\rho_{12}$ where $\rho_{12}$ is the correlation of the user portfolio and asset $s_2$, the equation \ref{eq:sr_appendix} can be re written as
\begin{equation} \label{eq:sr_appendix2}
  \frac{\mu_1|y_u| + \mu_2-r_f(|y_u|+1)}{\sqrt{\sigma_1^2{|y_u|}^2 + \sigma_2^2 + 2\sigma_{1}\sigma_{2}\rho_{12}|y_u|}}
\end{equation}

Please note that all the $w_1$, $w_2$, $\mu_1$, and $\sigma_1^2$ are given based on the user-item information, while $\mu_2$ and $\sigma_2^2$ are also ascertainable for any chosen asset $s_2$. The correlation between a user's portfolio and each item is computed using the weekly return series which is used when estimating the parameters $\mu$ and $\sigma^2$ as in Section 5.4. With the equation \ref{eq:sr_appendix2}, we can choose an item that can maximize the Sharpe ratio of an user's portfolio when the item is added to it, and we choose top $k$ items for each user to recommended as a baseline in order of the increased amount of Sharpe ratio.

\section{Hyperparameter Tuning}
Since MVECF is based on WMF, we need to consider the same set of hyperparameters of ordinary WMF. The hyperparameters of WMF are the latent dimension $l$, the regularization parameter ${\lambda}$, and the confidence level ${c_{ui}}$ of each observation ${y_{ui}} = 1$.

We performed a grid search within the following ranges $l \in \left\{ {10, 30, 50} \right\}$, ${c_{ui}} \in \left\{ {5, 10, 20, 40} \right\}$, and ${\lambda} \in \left\{0.0001, 0.001, 0.01\right\}$. The evaluation criteria for choosing the hyperparameter was MAP@20. The final hyperparameter values chosen for our experiments are $l = 30$, ${c_{ui}} = 10$ and ${\lambda} = $0.001. For SGD update of MVECF, we chose the learning rate ${\alpha}$ to 0.001, because the convergence was too slow for smaller learning rates and MVECF with large ${\lambda_{MV}}$ did not converge in some of our datasets for larger learning rates. 

\section{Train and Validation Loss}

\begin{figure}[h!]
\centering
\includegraphics[width=1\columnwidth]{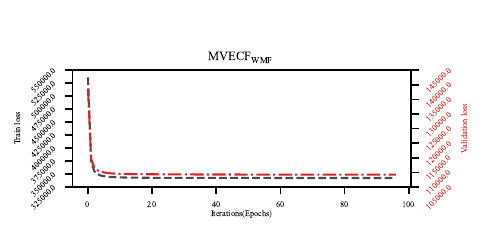}
\caption{Train and validation losses of $\text{MVECF}_\text{WMF}$.}
\label{fig:loss}
\end{figure}

$\text{BPR}_\text{nov}$ of Wasilewski \textit{et al.}~\shortcite{Wasilewski2019} is a modification of BPR model \cite{Rendle2009}. BPR uses matrix factorization to predict ratings by minimizing $\mathop \sum \nolimits_{u,i,j \in {\mathcal{D}_s}} - \log \sigma \left( {{{\hat y}_{ui}} - {{\hat y}_{uj}}} \right)$, where is the logistic sigmoid function, and ${\mathcal{D}_s}$ is a set of triples ($u, i,j)$ with ${y_{ui}} = 1$ and ${y_{uj}} = 0$.  Wasilewski \textit{et al.}~\shortcite{Wasilewski2019} defined a new set of triples $\mathcal{D}_s^{dist} = \{ \left( {u, i,j} \right):\left( {u, i,j} \right) \in {\mathcal{D}_s}{\text{\;and\;dist}}\left( {i,j} \right) < \tau\} $, where ${\text{dist}}\left( {i,j} \right)$ is the distance between items $i$ and $j$, and $\tau$ is a distance threshold. They train the model with $\left( {u,i,j} \right)$ sampled from $\mathcal{D}_s^{dist}$ with probability $\beta $ and sampled from ${\mathcal{D}_s}$ with probability $1 - \beta $ to recommend items that are distinct from the items in the user’s preference history. For BPR and $\text{BPR}_\text{nov}$, we set $l = 30$ and performed grid searches within the range ${\lambda} \in ${0, 0.00001, 0.0001, 0.001} and ${\alpha} \in \left\{ {0.001, 0.005, 0.01} \right\}$. The chosen values are ${\alpha} = 0.005$ and ${\lambda} = 0.00001$. We set the sampling probability $\beta$ to 80\% for $\text{BPR}_\text{nov}$,  and $\tau$ was set to 0.9 because it showed the best balance between novelty and recommendation performance among 0.8, 0.9, and 1.0. 
For graph based model, we used the same hyperparameters reported in paper and set the learning rate to 0.001, and the embedding size to 32. 
Figure \ref{fig:loss} shows the train loss and the validation loss of $\text{MVECF}_\text{wmf}$ in one of the yearly sub-datasets (year 2015) of CRSP data. We can easily see that both models converge quite smoothly. It implies that restructuring MVECF into an ordinary WMF form makes enhancement in the computational efficiency. Similar results are found in all other datasets.

\section{Sensitivity Analysis}

\begin{figure}[h!]
\centering
\includegraphics[width=1\columnwidth]{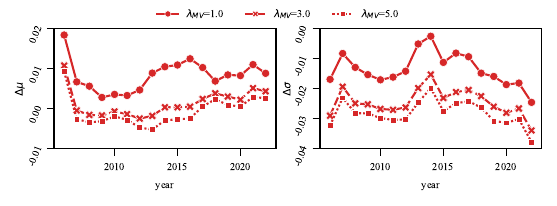}
\caption{$\Delta \mu$ and $\Delta \sigma$ of $\text{MVECF}_\text{wmf}$ with various values of $\gamma$.}
\label{fig:gamma_wmf}
\end{figure}

Figure \ref{fig:gamma_wmf} shows ${\Delta\mu}$ and ${\Delta\sigma}$ of $\text{MVECF}_\text{wmf}$ with ${\gamma} = \text{1, 3, and 5}$ for all our datasets when ${\lambda_{MV}}$ is set to 10. Note that ${\gamma}$ is a control parameter for determining how much to avoid risk, and it can be shown by the amount of changes in expected return ${\Delta\mu}$ and the changes in risk ${\Delta\sigma}$. We could find for all 20 sub-datasets, both ${\Delta\mu}$ and ${\Delta\sigma}$ decreased as ${\gamma}$ increases. As larger ${\gamma}$ implies more risk-averse user, MVECF focuses more on reducing risk than increasing return. Hence, ${\gamma}$ can control the risk-aversion as desired. It is worth noting that $\text{MVECF}_\text{wmf}$ effectively increases return (positive ${\Delta\mu}$) when ${\gamma} = 1$. The results show that $\text{MVECF}_\text{wmf}$ effectively controls both the recommendation-investment performance trade-off and the risk-return trade-off of the portfolios.

\begin{figure*}[t!]
\centering
\includegraphics[width=2\columnwidth]{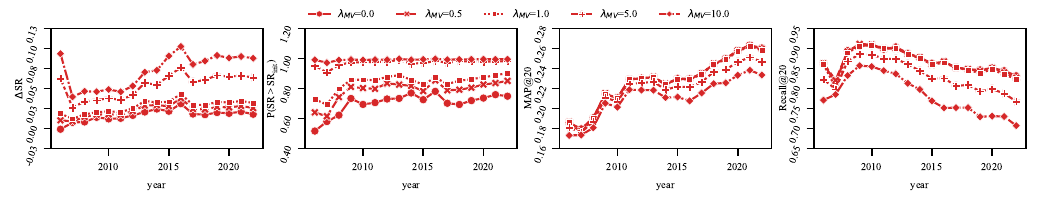}
\caption{Performance of $\text{MVECF}_\text{wmf}$ with various values of $\lambda_{MV}$.}
\label{fig:lambda_mv_wmf}
\end{figure*}

Figures \ref{fig:lambda_mv_wmf} shows the evaluation metrics of $\text{MVECF}_\text{wmf}$ with various values of ${\lambda_{MV}}$ for a total of 17 sub-datasets (17 yearly sub-datasets for CRSP data, 10 yearly sub-datasets for Thomson Reuters data). Each marker in the graph represents the average performance of all users in one yearly sub-dataset. The risk-aversion parameter ${\gamma}$ is fixed as 3. Note that when ${\lambda_{MV}} = 0$, MVECF becomes an ordinary WMF model. The figures show that the MV efficiency of MVECF model is enhanced as $\lambda_{MV}$ increases in all our datasets. In addition, MAP@20 and Recall@20 decreases as $\lambda_{MV}$ varies from 0 to 10. It shows the trade-off relationship between the MV efficiency and conventional recommendation performance which indicate that MVECF can control the balance between the user’s preference and the MV efficiency by varying the MV regularization level.

\vfill\eject